\RequirePackage{fix-cm}
\documentclass[smallextended]{svjour3}       
\smartqed  
\usepackage{graphicx}
\usepackage{miller}
\begin{document}

\title{Beam model for the elastic properties of material with spherical voids}


\author{Sascha Heitkam       \and        Wiebke Drenckhan \and        Denis Weaire     \and        Jochen Fr\"ohlich
}


\institute{S. Heitkam, J. Fr\"ohlich \at
              Institute of Fluid Mechanics, Technische Universit\"at Dresden, 01069 Dresden, Germany  \\
              Tel.: +49-351-463 34910\\
              Fax: +49-351-463 35246\\
              \email{sascha.heitkam@tu-dresden.de}           
           \and
           W. Drenckhan \at
              Laboratoire de Physique des Solides, Universit\'e Paris Sud XI, 91405 Orsay, France
                         \and
           D. Weaire \at
              School of Physics, Trinity College Dublin 2, Ireland
}

\date{Received: date / Accepted: date}

\maketitle

\begin{abstract}

The elastic properties of a material with spherical voids of equal volume are analysed using a new model, with particular attention paid to the hexagonal close-packed and the face-centred cubic arrangement of voids.
Void fractions well above 74 \% are considered, yielding overlapping voids as in an open-cell foam and hence a connected pore structure. The material is represented by a network of beams. The elastic behaviour of each beam is derived analytically from the material structure. By computing the linear elastic properties of the beam network, the Young's moduli and Poisson ratios for different directions are evaluated. In the limit of rigidity loss a power law is obtained, describing the relation between Young's modulus and void fraction with an exponent of 5/2 for bending-dominated and 3/2 for stretching-dominated directions. The corresponding Poisson ratios vary between 0 and 1. With decreasing void fraction, these exponents become 2.3 and 1.3, respectively. The data obtained analytically and from the new beam model are compared to Finite Element simulations carried out in a companion study, and good agreement is found. The hexagonal close-packed void arrangement features very anisotropic behaviour, comparable to that of fibre-reinforced materials, This might allow for new applications of open-cell materials.  

\end{abstract}

\section{Introduction}

\begin{figure}
\centering
  \includegraphics[width=0.65\textwidth]{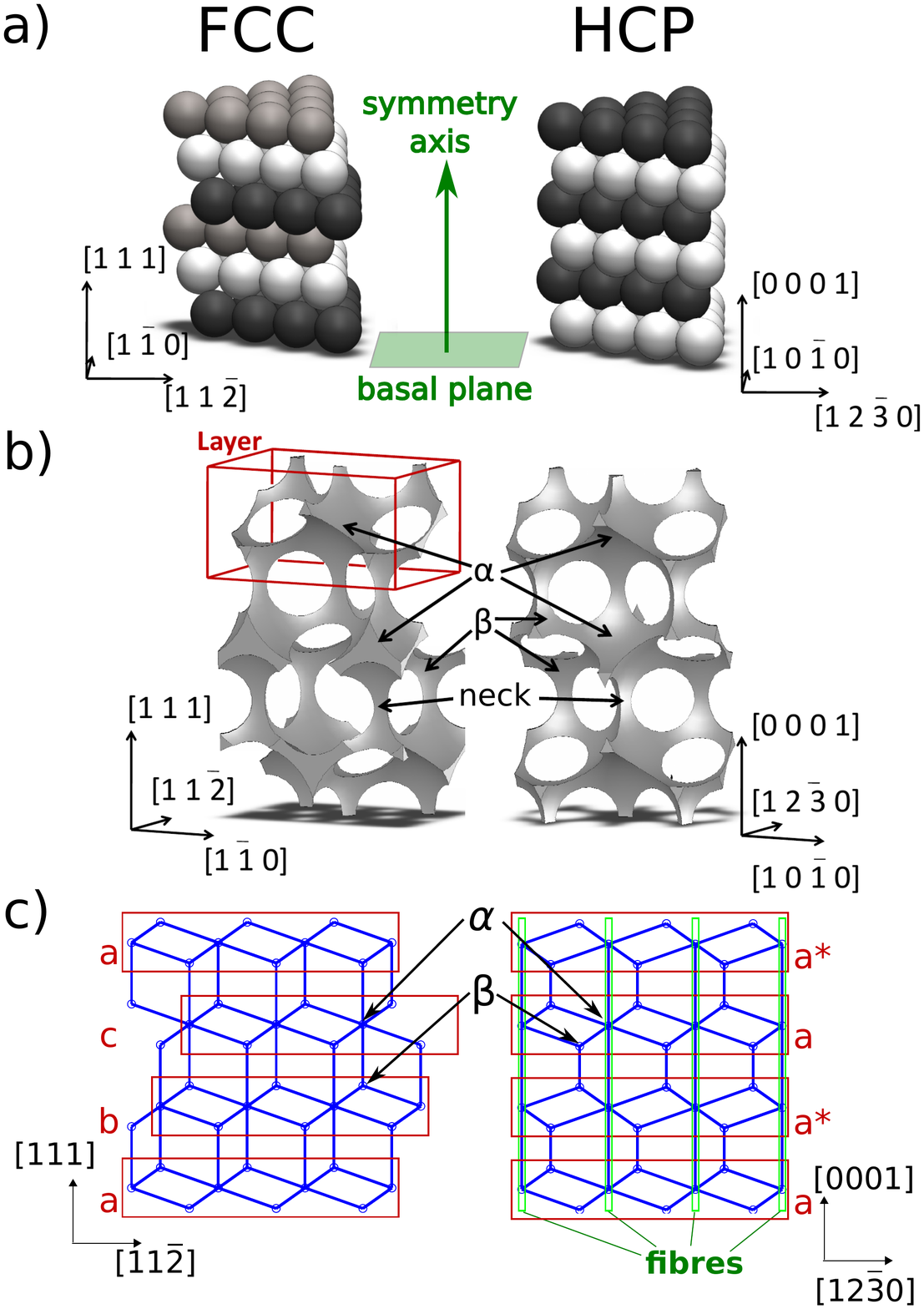}
\caption{Structure of void material. (a) Arrangement of spheres in face-centred cubic ordering (left) and hexagonal close-packed ordering (right). A particular symmetry axis corresponds to the \hkl [1 1 1] and \hkl [0 0 0 1] direction, respectively, and is oriented normally to the basal plane. (b) Structure of the resulting void material displaying nodes $\alpha$ and $\beta$ and necks. One layer is indicated, which may be stacked to build up FCC or HCP void material. (c) The corresponding beam model consisting of vertices $\alpha$ and $\beta$ connected by straight beams. The arrangement of material layers and the existence of fibres are highlighted.}
\label{fig:void_material}       
\end{figure}

In the present work the elastic properties of void material are investigated. This is a homogeneous material such as a polymer or metal, filled with spherical voids of uniform diameter, arranged on a crystalline lattice. These voids overlap, forming an open-cell structure, as shown in Figure~\ref{fig:void_material}b for a void fraction of 85\%. Such a structure can be manufactured, for example, using sacrificial templating~\cite{Kikuchi2011} or 3D-printing techniques. Open-cell polymer foams have similar geometries, but in that case surface tension effects assure a more even distribution of the solid material in the network~\cite{Mills2007,Warren1997}.\\
Such a material offers particular and very interesting properties. Currently, void materials are often used in lightweight construction, cushioning or shock absorption, although usually these are disordered foams.
\\In sacrificial templating and 3D-printing it is possible to arrange the voids in a specified regular lattice. For metal foams, application of an electromagnetic field also promotes specific bubble arrangement to some extent~\cite{Heitkam2013}. Depending on the arrangement of voids, specialised anisotropic elastic properties can result. In particular, these properties could give rise to anisotropic acoustic wave propagation, opening up new applications.\\
To analyse the elastic properties of these materials, highly resolved Finite Element (FE) simulations were recently carried out by the authors, yielding the full stiffness matrix with an uncertainty lower than $1 \%$. These simulations are documented in detail in a companion study~\cite{Heitkam2015}. However, it is difficult to gain insight directly from these simulations, as regards the key mechanisms that lead to dependence of elastic properties on topology and void fraction of the material. Also, for very high void fractions the interstitial material becomes very thin and FE simulations are hardly applicable due to problems of grid generation.\\
Therefore, in the present study a model composed of elastic beams is applied which is based on a skeletal representation of the topology and geometry of the void material. In the past, Truss models have been used to compute the mechanical properties of shells~\cite{Mbakogu1987}, plates~\cite{Salonen1971} with low computational effort. There, homogeneous material is represented by a network of beams. In the present work, heterogeneous void material is considered and each beam reflects a structural element.  
The results of the beam model are validated using our previous FE simulations~\cite{Heitkam2015}. Most of all, the beam model allows for a direct study of the necessary ingredients which are at the origin of the observed elastic behaviour, revealing the underlying mechanisms. Taking into account analytical considerations, the elastic properties of void material with very high void fractions is estimated. This approach is motivated by the work of Day et al.~\cite{Day1992}. These authors computed the elastic properties of two-dimensional sheets with circular holes using an FE method and compared these results with analytic approximations taking into account only the deformations around the weakest regions.\\
In three dimensions, one can find several studies investigating the elastic properties of open-cell materials by modelling them with networks of beams~\cite{Mills2007,Gibson1997,Warren1997,Roberts2002}. In that way, one can estimate Young's modulus and Poisson ratio of the given topology. The best known among these are the results of Gibson and Ashby~\cite{Gibson1997}. These authors predict the behaviour of low density foam by analysing an artificial beam structure with and without thin lamellas between the voids. The beams usually are taken to have square cross sections. Such a representation is also adopted in the present work. But the shape, topology and elastic constants of the beams are extracted to a higher level of accuracy from the interstitial material, leading to substantially different results, as described below. 

\section{Beam model}
\subsection{Void structures}
In this work, the considered crystalline arrangements of voids are face-centred cubic (FCC) and hexagonal close-packed (HCP) which are of very similar structure. Both arrangements consist of layers of hexagonally packed spheres, but these layers are stacked in a different sequence, as demonstrated in Figure~\ref{fig:void_material}a. The numbers in brackets refer to the Miller-Bravais indices.
 In case of FCC, the sequence is a-b-c-a-b-c, where a, b and c correspond to different relative positions of the layers. In case of HCP the sequence is a-b-a-b. Both structures achieve the densest packing of spheres which has 74\% void fraction~\cite{Weaire2008}. The spherical voids used here have diameters which are larger those corresponding to close packing, which gives rise to overlapping spheres in which the overlapping material is removed from the structure.\\
The interstitial material of both sphere packings consists of layers, confined between two layers of spherical voids as marked in Figure~\ref{fig:void_material}b. In the FCC and HCP void arrangement, the same type of material layer is formed, but again, these material layers are stacked in different sequence. In FCC, the sequence is a-b-c-a-b-c, similar to the sequence of sphere layers. In HCP, the sequence is a-a*-a-a*, where a* denotes a flipped material layer. The arrangement of the layers is indicated in Figure~\ref{fig:void_material}c.\\

\subsection{Beam network}
\begin{figure}
  \centering
	  \includegraphics[width=1.00\textwidth]{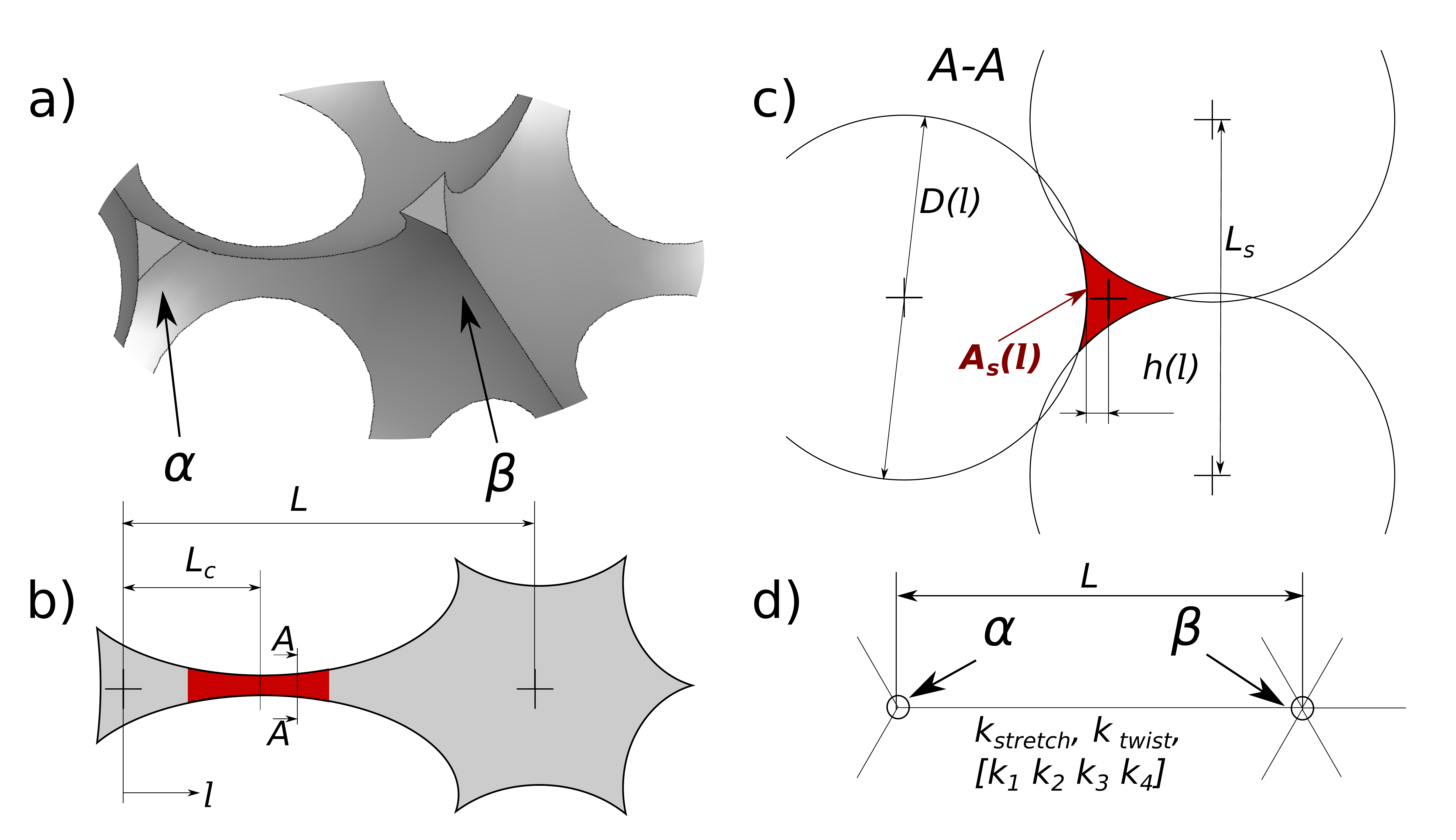}
	  \caption{(a) Original geometry of necks and nodes. (b) Identification of the narrow part of the neck, which is highly deformable and dominates elastic properties. (c) Geometry of the cross section of a neck. (d) Reduction to a thin beam connecting two vertices with specific elastic constants.}
    \label{fig:beam}       
\end{figure}

Let us now focus on the interstitial material which forms a network of nodes and necks (Figure~\ref{fig:void_material}b). Comparing FCC and HCP, both networks consist of the same elements but show a different topology. There are two types of nodes, $\alpha$ and $\beta$, displayed in Figure~\ref{fig:void_material}b. The large node $\alpha$ is confined between 6 voids in octahedral arrangement and connected to 8 necks. The small node $\beta$ is confined between 4 voids in tetrahedral arrangement and connected to 4 necks. Each neck connects two nodes. The critical, most deformable region of each neck has essentially the same shape (Figure~\ref{fig:beam}) and is confined between three spherical voids.
The next step is to develop an approximate representation of the material in terms of thin straight beams connected by point vertices (Figure~\ref{fig:void_material}c and Figure~\ref{fig:beam}c). Each beam corresponds to one neck and each vertex to one node. The total length $L$ of a beam equals the separation of the node centres and depends on the type of vertices it connects.
Each point vertex is treated as follows: a balance of forces and torques is sought, determining its translation and rotation. This causes the attached beams to deform. In turn, the deformation of beams generates forces and torques on the attached vertices. In an iterative scheme, the positions and angles of the vertices, corresponding to a minimum elastic energy of the network, are determined.
\\ In order to compute the linear elastic properties of the void material, a periodic unit cell is extracted and represented as a beam network. Periodic boundaries are applied by communicating force and position between periodic vertex pairs. In one direction, e.g. the \hkl [1 1 1] direction (Miller-Bravais indices), a small strain $\epsilon(111)$ is applied. After relaxation of the beam network the resulting stress $\sigma(111)$ is evaluated. In the transversal directions, \hkl [1 -1 0] and \hkl [1 1 -2], zero-stress boundary conditions are applied and the strain $\epsilon(1\bar{1}0)$ and $\epsilon(11\bar{2})$ is evaluated as well. Subsequently, the relation of stress and strain yields the Young's modulus and the Poisson ratio of the void material for the selected directions, which in this case are
\begin{eqnarray}
 E(111) &=& \sigma(111) / \epsilon(111) \\
 \nu(111,1\bar{1}0) &=& -\epsilon(1\bar{1}0) / \epsilon(111),
\end{eqnarray}
respectively.

\subsection{Neck geometry}
\begin{figure}
    \centering
	\includegraphics[width=0.75\textwidth]{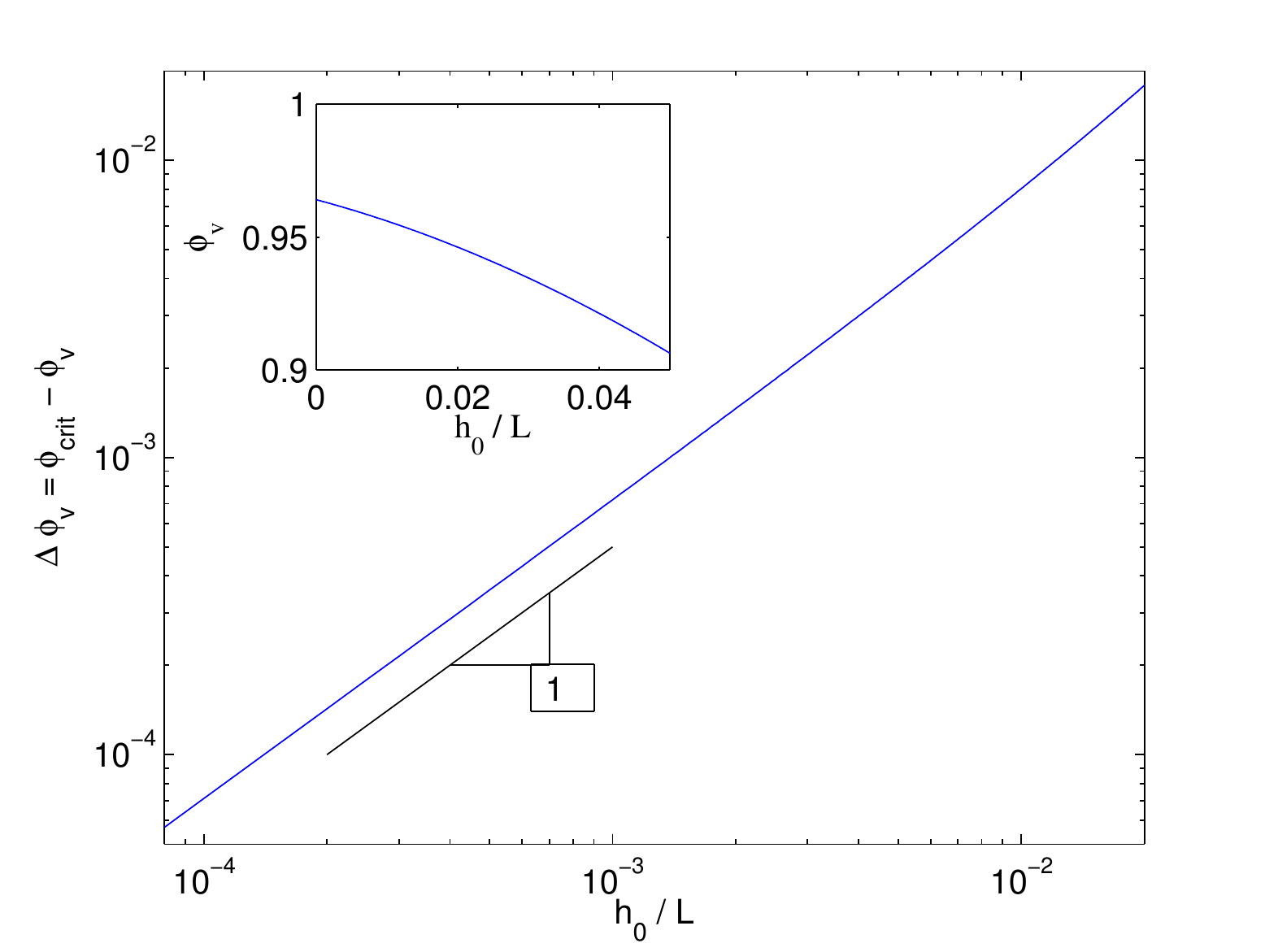}
	\caption{Relation between beam thickness $h_0 = h(l=L_c)$ and void fraction $\phi_v$, derived from analytical calculations of geometrical relations.}
    \label{fig:phi_h}       
\end{figure}

In earlier work~\cite{Mills2007,Gibson1997,Warren1997,Roberts2002} it was found convenient tu use the void fraction $\phi_v$ as a parameter for the elastic behaviour. Here, the void fraction is varied by changing the diameter $D_0$ of the voids without changing their centre spacing $L_s$. In that way, the void fraction changes the effective thickness $h (l)$ of the beams, their cross-sectional area $A_s (l)$, and therefore their elastic properties.\\
The position coordinate along the neck is denoted by $l$ (Figure~\ref{fig:beam}b). The most slender cross section of the neck is situated at $l=L_c$ and is confined by three circles with diameter $D_0$ (Figure~\ref{fig:beam}d). At this position the beam thickness $h$ attains its minimum $h=h_0$, with $h$ being defined as the distance between the centroid of the cross section and the closest point on the surface of the cross section (Figure~\ref{fig:beam}d). With increasing distance $\tilde{l} = \left| l-L_c \right|$ from the point of minimal beam thickness, $h(\tilde{l})$ and the cross-sectional area $A_s (\tilde{l})$ increase according to
\begin{eqnarray}
 h(\tilde{l}) &=& h_0 + \frac{D_0}{2} -\sqrt{ \frac{D_0^2}{4}-\tilde{l}^2} \approx h_0 + \frac{\tilde{l}^2}{D_0} + o(\tilde{l}^4) , \label{eq:hl}\\
 A_s(\tilde{l}) &\sim& h(\tilde{l})^2 + o(\tilde{l}^4). \label{eq:Al}
\end{eqnarray}
 Figure~\ref{fig:phi_h} provides the relation between void fraction $\phi_v$ and beam thickness $h_0$ resulting from tedious geometrical evaluations. For a critical void fraction $\phi_v = \phi_{crit}$ the minimal beam thickness becomes zero. This means that the beams are cut through and the structure is not connected any more. Below this critical value the beam thickness increases linearly with the difference $\Delta \phi_v = \phi_{crit}-\phi_v$ which is a good approximation over a wide range of beam thickness.

\section{Beam properties}
\begin{figure}
\centering
  \includegraphics[trim = 30mm 50mm 50mm 50mm, clip, width=0.7\textwidth]{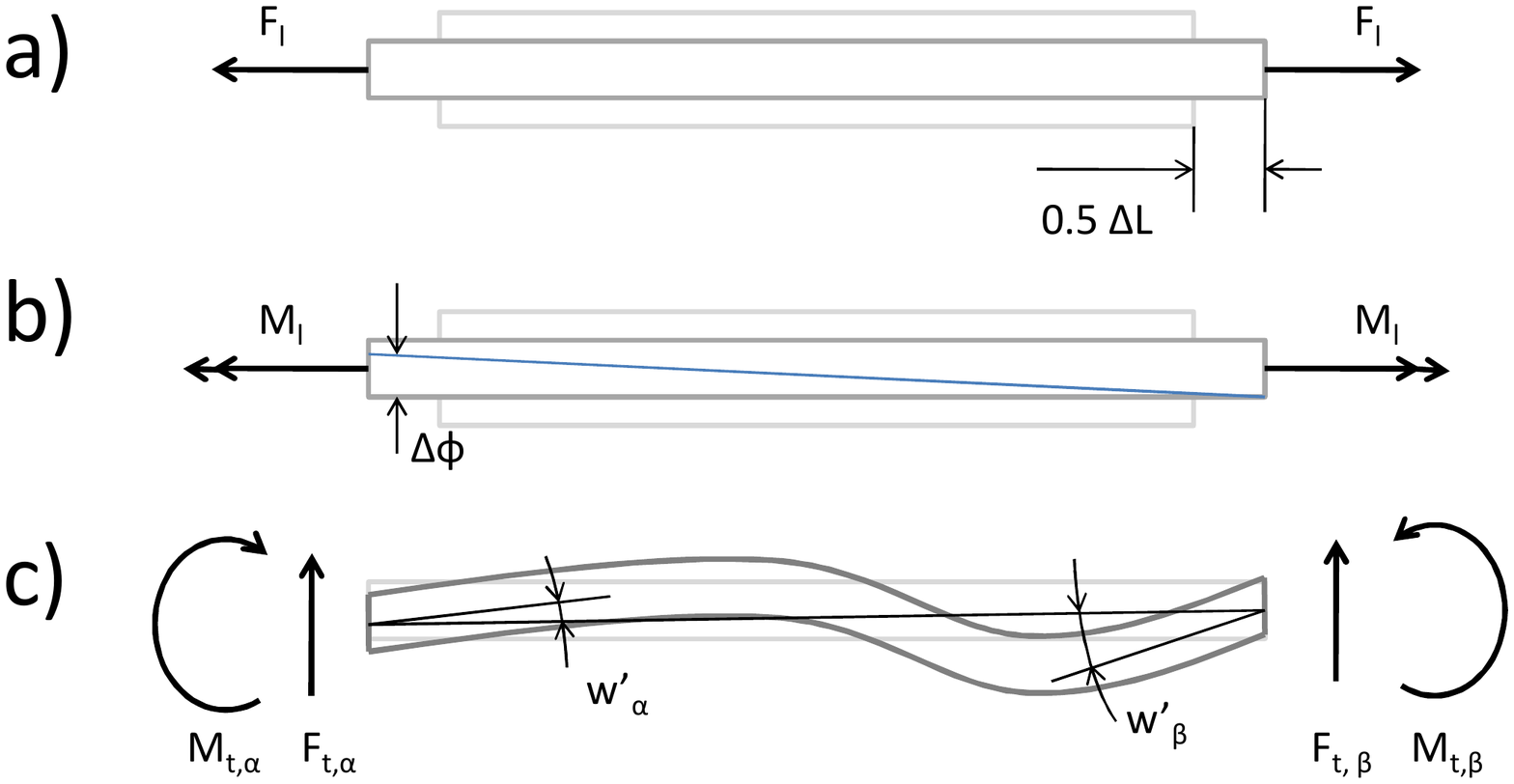}
\caption{Possible deformations for a beam. (a) Stretching by $\Delta L$ caused by a longitudinal force $F_l$. (b) Twisting by an angle $\Delta \Phi$, caused by a longitudinal torque $M_l$. (c) Bending, caused by transversal forces $F_t$ and torque $M_t$, applied at the vertices $\alpha$ and $\beta$. Bending is quantified by a mismatch between conjugation line of the attached vertices and first derivative of the beam deflection $w'$ at the vertices. For illustration a rectangular beam is shown. The actual neck cross section varies along the length.}
\label{fig:deform}       
\end{figure}

For each beam, three principal kinds of deformation are defined: stretching, twisting and bending. These are illustrated in Figure~\ref{fig:deform} using the example of beams with homogeneous thickness.
In case of small deformation, one may assume these three types of deformation to be superimposed without interference. The position and angle of the adjoined vertices imposes a deformation on the beam. With given deformation, beam theory allows us to compute for each beam the corresponding reaction forces applied to the vertices. To do so, one has to define for each kind of deformation an elastic constant ($k_{\mathrm{stretch}}$, $k_{\mathrm{twist}}$, $k_{\mathrm{bend}}$) for a beam, which will be done in the following subsections.

\subsection{Stretching}
Stretching refers to the change $\Delta L$ of the length $L$ of a beam when applying a longitudinal force $F_l$. This response is calculated by integrating the incremental strain along a neck, i.e. 
\begin{eqnarray}
 \frac{\Delta L }{ L} &=& \frac{F_l}{L E_0} \int_0^L \frac{1}{A_s(l)} dl \\ \label{eq:el1}
 \frac{\Delta L }{ L} &=& \frac{F_l}{L E_0} \int_{-\infty}^{+\infty} \frac{1}{A_s(l)} dl = F_l \frac{1}{k_{\mathrm{stretch}}}.
  \label{eq:el2}
\end{eqnarray}
The elastic constant $k_{\mathrm{stretch}}$ depends on the Young's modulus $E_0$ of the solid material and the cross-sectional area $A_s$. Since the area $A_s$ is much smaller in the critical region of a neck, the integrand in~(\ref{eq:el1}) falls off rapidly and hence becomes negligible elsewhere. Consequently, the integration can be carried out in the bounds of $[ -\infty , \infty]$ in Equation~(\ref{eq:el2}) without changing the resulting value. This identity is important to derive an analytical scaling law for the elastic constants in Section~\ref{sec:scaling} below without tedious consideration of boundaries. 

\subsection{Twisting}
Applying longitudinal torque $M_l$ to a beam causes a change of the longitudinal angle  
\begin{equation}
   \Delta \Phi = \frac{M_l}{G_0} \int_{-\infty}^{+\infty} \frac{1}{I_p(l)} dl = \frac{M_l }{L} \frac{1}{k_{\mathrm{twist}}}.
   \label{eq:tw1}
\end{equation}
To determine the elastic constant $k_{\mathrm{twist}}$  one has to integrate the polar second moment of area $I_p (l)$. Due to the complex geometry, $I_p$ was approximated by $I_p (l) = A_s (l)  h^2 (l)$. The shear modulus $G_0 = E_0 (2(1+\nu_0))^{-1}$ results for the isotropic solid material from the given Poisson ratio $\nu_0$ and Young's modulus $E_0$.  

\subsection{Bending}
Bending of a beam causes transversal torque $M_t$ and force $F_t$ on the adjoined vertices $\alpha$ and $\beta$. They depend on the first derivative $w' = dx_t / dl$ of the deflection close to the adjoined vertices $w'_{\alpha}$ and $w'_{\beta}$. First, four elastic constants $k_{1 \ldots 4}$ have to be derived
\begin{eqnarray}
 k_1 &=& \frac{1}{E_0} \int_{-\infty}^{+\infty} \frac{1}{I_t(l)} dl \\
 k_2 &=& \frac{1}{E_0} \int_{-\infty}^{+\infty} \frac{l}{I_t(l)} dl \\
 k_3 &=& \frac{1}{E_0} \int_{0}^{L} \int_{0}^{\tilde{x}} \frac{1}{I_t(l)} dl d\tilde{x} \\
 k_4 &=& \frac{1}{E_0} \int_{0}^{L} \int_{0}^{\tilde{x}} \frac{l}{I_t(l)} dl d\tilde{x}.
\end{eqnarray}
Here the bounds $0$ and $L$ are retained because the deflection of the neck may only be taken into account between the nodes. In contrast to stretching and twisting, the result depends on the distance between the nodes and is not the same for all necks. The second moment of area depends on the shape of the cross section and was approximated by $I_t (l) = 0.5 A_s (l) h^2 (l)$. This expression does not contain any dependence on the transversal direction, so that the elasticity of beams is assumed to be isotropic in transversal direction.
Subsequently, force and torque can be computed by
\begin{eqnarray}
 F_{t,\alpha}&=&\left( k_2 - \frac{k_1 k_4}{k_3}\right)^{-1} \left[ w'_{\beta} + \left(\frac{k_1}{k_3} L -1 \right) w'_{\alpha} \right] \\
 M_{t,\alpha}&=&\left( k_1 - \frac{k_2 k_3}{k_4}\right)^{-1} \left[ w'_{\beta} + \left(\frac{k_2}{k_4} L -1 \right) w'_{\alpha} \right] \\
 F_{t,\beta}&=& - F_{t,\alpha} \\
 M_{t,\beta}&=& M_{t,\alpha} + F_{t,\alpha} L \label{eq:lab1}\\
 k_{bend} &=& \left( k_2 - \frac{k_1 k_4}{k_3}\right)^{-1} \left[ 1 + \left(\frac{k_1}{k_3} L -1 \right) \right].
\end{eqnarray}
The required values of $w'_{\alpha}$ and $w'_{\beta}$ can be derived from the mismatch between angular orientation and the line connecting the adjoined vertices (Figure~\ref{fig:deform}c). The simple bending constant $k_{bend}$, linking force $F_t$ and tilt $w'$, is introduced here to construct a viable iterative scheme and to compute scaling behaviour below.

\subsection{Scaling of effective beam constants}
\label{sec:scaling}
\begin{figure}
\centering
  \includegraphics[width=0.75\textwidth]{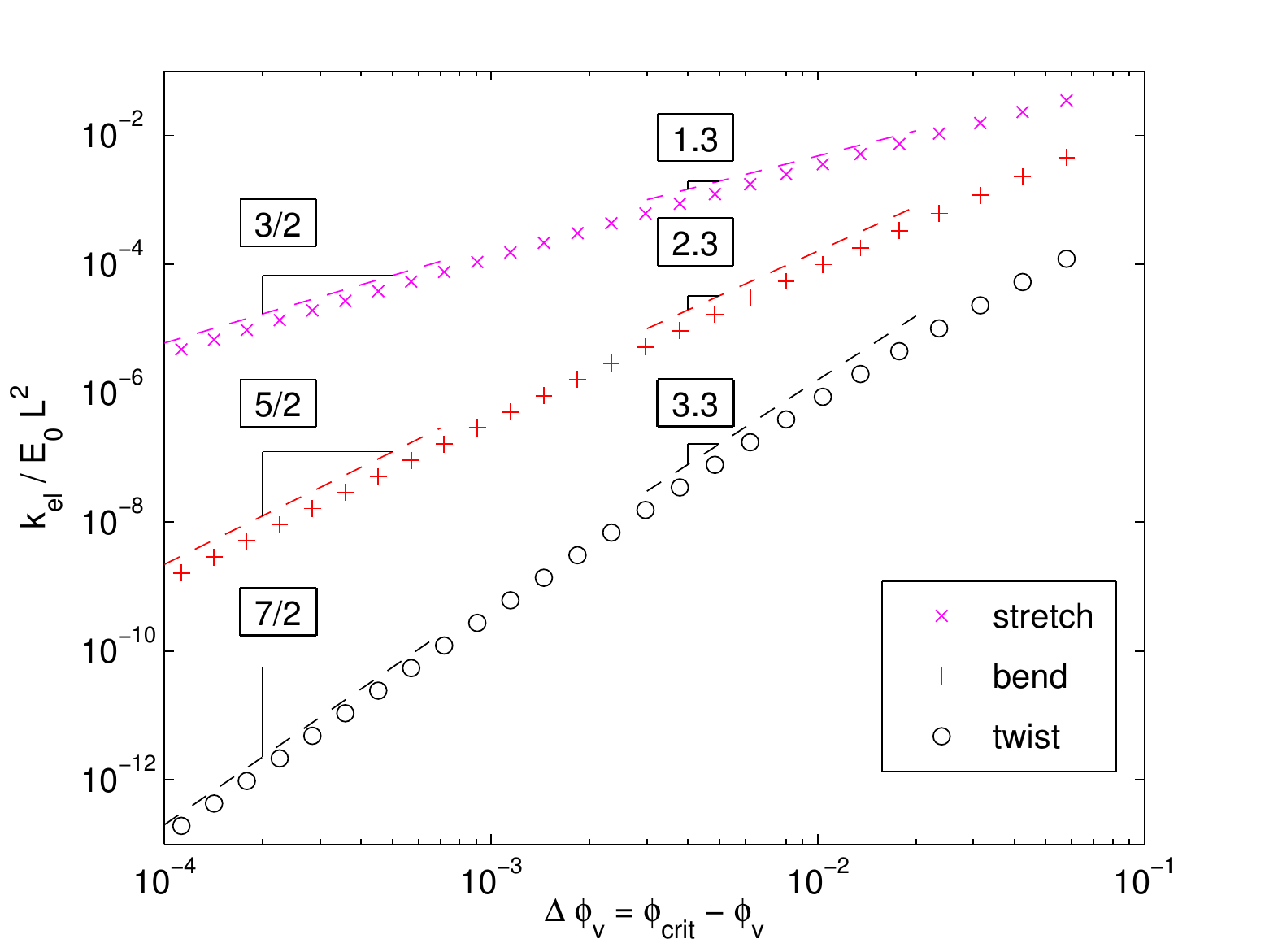}
\caption{Dependence of the effective beam constants on the void fraction difference $\Delta \phi_v = \phi_{crit}-\phi_v$. For the limit of rigidity loss a power law with exponents $3/2$ for stretching, $5/2$ for bending and $7/2$ for twisting is valid. For higher void fraction difference the exponents decrease to $1.3$, $2.3$ and $3,3$.}
\label{fig:k_el}       
\end{figure}

If the void fraction equals the critical void fraction, the void material becomes disconnected and the Young's modulus equals zero. With decreasing void fraction the beam thickness $h_0$ and thus, the Young's modulus increases. The relation between void fraction and the elastic constants $k_{stretch}$, $k_{bend}$ and $k_{twist}$ is developed in a power law in terms of the difference $\Delta \phi_v = \phi_{crit}-\phi_v$.
Applying the approach of Day et al.~\cite{Day1992} step-by-step reveals an analytical derivation of the exponents of this scaling.\\
The void fraction difference $\Delta \phi_v \sim h_0$ scales linearly with the beam thickness $h_0$ (Figure~\ref{fig:phi_h}). The thickness of the beam along the neck can be developed in a Taylor series yielding $h(\tilde{l}$ and $A_s(\tilde{l})$ according to (\ref{eq:hl}) and (\ref{eq:Al}).
 The cross section area $A_s (l) \sim h^2(l)$ scales quadratically with the thickness $h(l)$. Inserting these into Equation~(\ref{eq:el2}) yields
 \begin{eqnarray}
 \frac{1}{k_{stretch}} &\sim& \int_{-\infty}^{+\infty} \frac{1}{A_s (l)} dl \sim \int_{-\infty}^{+\infty} \frac{1}{h^2 (l)} dl = \frac{\pi \sqrt{D_0}}{2  } \frac{1}{h_0^{3/2}} .
\end{eqnarray}
Consequently, the stretching stiffness $k_{stretch}$ scales as
\begin{equation}
  k_{stretch} \sim h_0^{3/2} \sim (\Delta \phi_v)^{3/2} 
\end{equation}
Analogously, the exponents for $k_{bend}$ and $k_{twist}$ can be shown to equal $5/2$ and $7/2$, respectively 
 \begin{eqnarray}
  \frac{1}{k_{bend}} &\sim& \left( k_2 - \frac{k_1 k_4}{k_3} \right)       \sim \sqrt{D_0} \frac{ 1}{h_0^{5/2} } \label{eq:kb}\\
   \frac{1}{k_{twist}} &\sim& \int_{-\infty}^{+\infty} \frac{1}{h^2(l) A_s (l)} dl \sim \int_{-\infty}^{+\infty} \frac{1}{h^4 (l)} dl = \frac{15 \pi \sqrt{D_0}}{48  } \frac{1}{h_0^{7/2}}.
\end{eqnarray}
For $k_{bend}$ the parameter $k_1, \ldots, k_4$ scale with $h_0^{-7/2}$ but due to the subtraction in Equation~(\ref{eq:kb}) the leading term vanishes and the second largest term $\sim h_0^{-5/2}$ dominates the scaling.\\
In conclusion, the following scaling laws for the elastic properties of a beam are derived in the limit of vanishing beam thickness, i.e. $h_0 \rightarrow 0$ and $\Delta \phi_v \rightarrow 0$
\begin{eqnarray}
  k_{stretch} &= \frac{F_l L}{\Delta L} &\sim (\Delta \phi_v)^{3/2} \\
  k_{bend} &= \frac{F_t}{w'_{\alpha}} &\sim (\Delta \phi_v)^{5/2}\\
  k_{twist} &= \frac{M_l}{\Delta \Phi L}  &\sim (\Delta \phi_v)^{7/2}. 
  \end{eqnarray}
For values of $h_0$ and $\delta \phi_v$ remote from this limit addressed analytically, the elastic properties were computed numerically according to the approximations described above in~(\ref{eq:el1})~-~(\ref{eq:lab1}). The results are shown in Figure~\ref{fig:k_el}
over a wide range of void fractions. For void fractions $0.003 \leq \Delta \phi_v \leq 0.03$  Figure~\ref{fig:k_el} suggests slightly lower exponents in the scaling of the elastic constants, namely
\begin{eqnarray}
  k_{stretch} &= \frac{F_l L}{\Delta L} &\sim (\Delta \phi_v)^{1.3} \\
  k_{bend} &= \frac{F_t}{w'_{\alpha}} &\sim (\Delta \phi_v)^{2.3} \\
    k_{twist} &= \frac{M_l}{\Delta \Phi L}  &\sim (\Delta \phi_v)^{3.3} .
\end{eqnarray}
which were obtained by linear regression.

\section{Results}
The elastic behaviour of void material may be characterised by its Young's modulus and Poisson ratio. 
\subsection{Young's modulus}
\begin{figure}
\centering
  \includegraphics[width=0.75\textwidth]{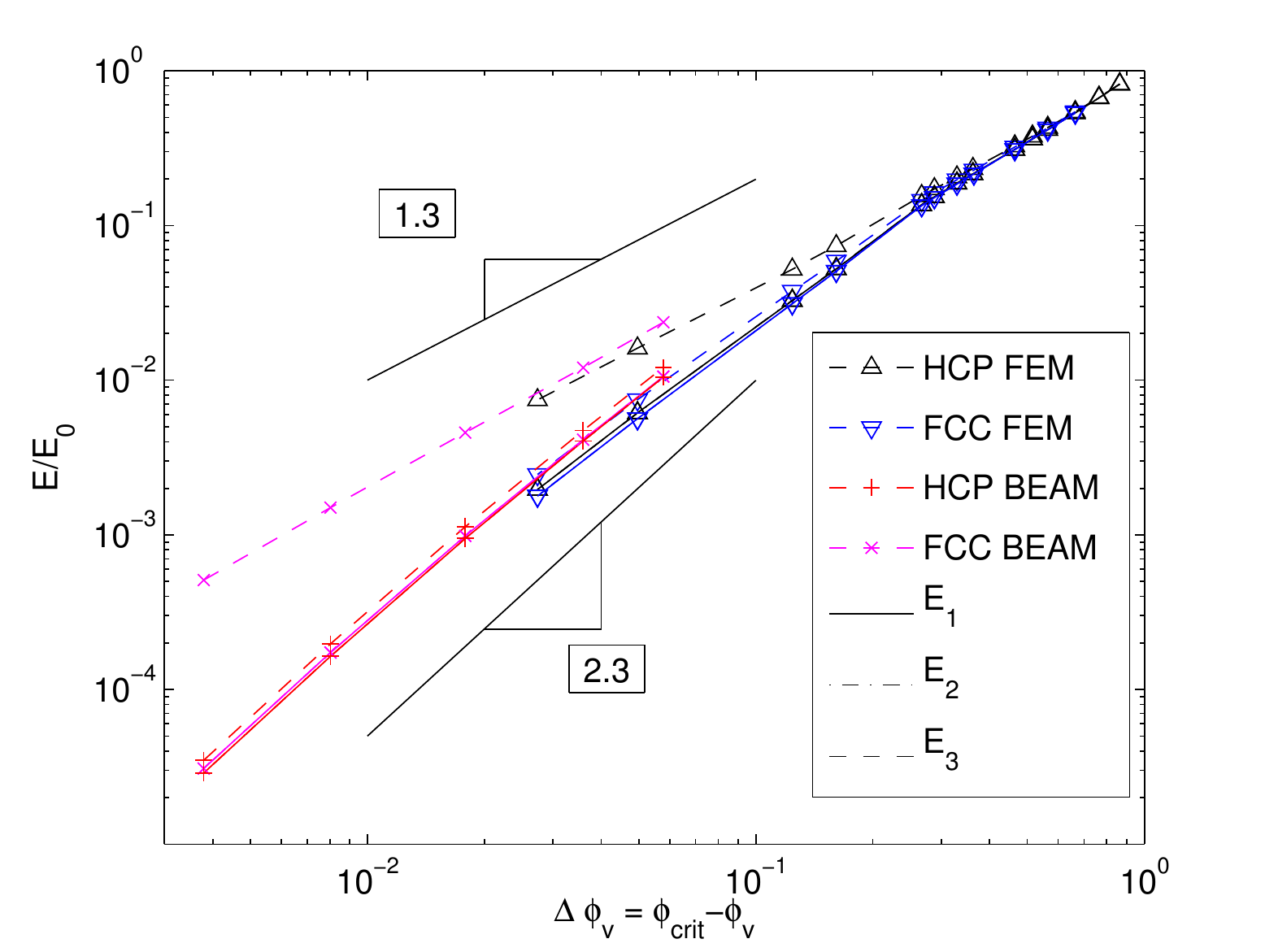}
\caption{Young's modulus as a function of the difference between void fraction and critical void fraction. Data of FE simulations ($\nabla$, $\Delta$) and Beam model ($\times$,+) for FCC ($\nabla$, $\times$) and HCP ($\Delta$,+). Line style indicates the direction: $E_3$ (- -) corresponds to the symmetric direction; $E_1$ (-.) and $E_2$ (-) corresponds to directions in the basal plane. Due to isotropy in the basal plane, $E_1$ and $E_2$ coincide.  }
\label{fig:E}       
\end{figure}

In Figure~\ref{fig:E} the Young's modulus is plotted against the distance $\Delta \phi_v$ between void fraction and critical void fraction, which according to Figure~\ref{fig:phi_h} is related to the beam thickness. Figure~\ref{fig:E} shows the results obtained from the beam model and those obtained from the FE simulations of FCC and HCP void arrangement in~\cite{Heitkam2015}.
Even though small deviations are visible, the basic dependences and relations are very well reproduced by the beam model. Solid and dash-dotted lines correspond to directions in the basal plane. In the figure, only the solid lines are visible, because the graphs coincide nearly perfectly, revealing isotropy of the Young's modulus in the basal plane. This finding is also in line with the results of the companion FE simulations. The Young's modulus of HCP in the basal plane is slightly higher. Comparing the symmetric axis, larger differences are visible. For FCC, the Young's modulus is in the same range as the other ones, but for HCP the Young's modulus in the direction of the symmetric axis is much higher. Also the dependence on the void fraction is different. For HCP in symmetric direction the Young's modulus is proportional to the void fraction to the power of 1.3 while the Young's modulus in the other directions and for FCC are proportional to the void fraction to the power of 2.3. The reason is the straight uninterrupted connection of beams in the HCP void material, called fibre here. These fibres transfer load along the symmetric axis by stretching, and $k_{stretch}$ increases with the void fraction to the power of 1.3, as shown in Figure~\ref{fig:k_el}. In the other directions and for FCC, the Young's modulus increases with the void fraction to the power of 2.3. For these cases force transfer always causes bending of beams (see Figure~\ref{fig:void_material}) and for the relevant $k_{bend}$ the same power law holds. Consequently, the stiffness of HCP void material decreases more slowly than that of FCC when approaching the critical void fraction. 

\subsection{Poisson ratio}
\begin{figure}
\centering
  \includegraphics[width=0.75\textwidth]{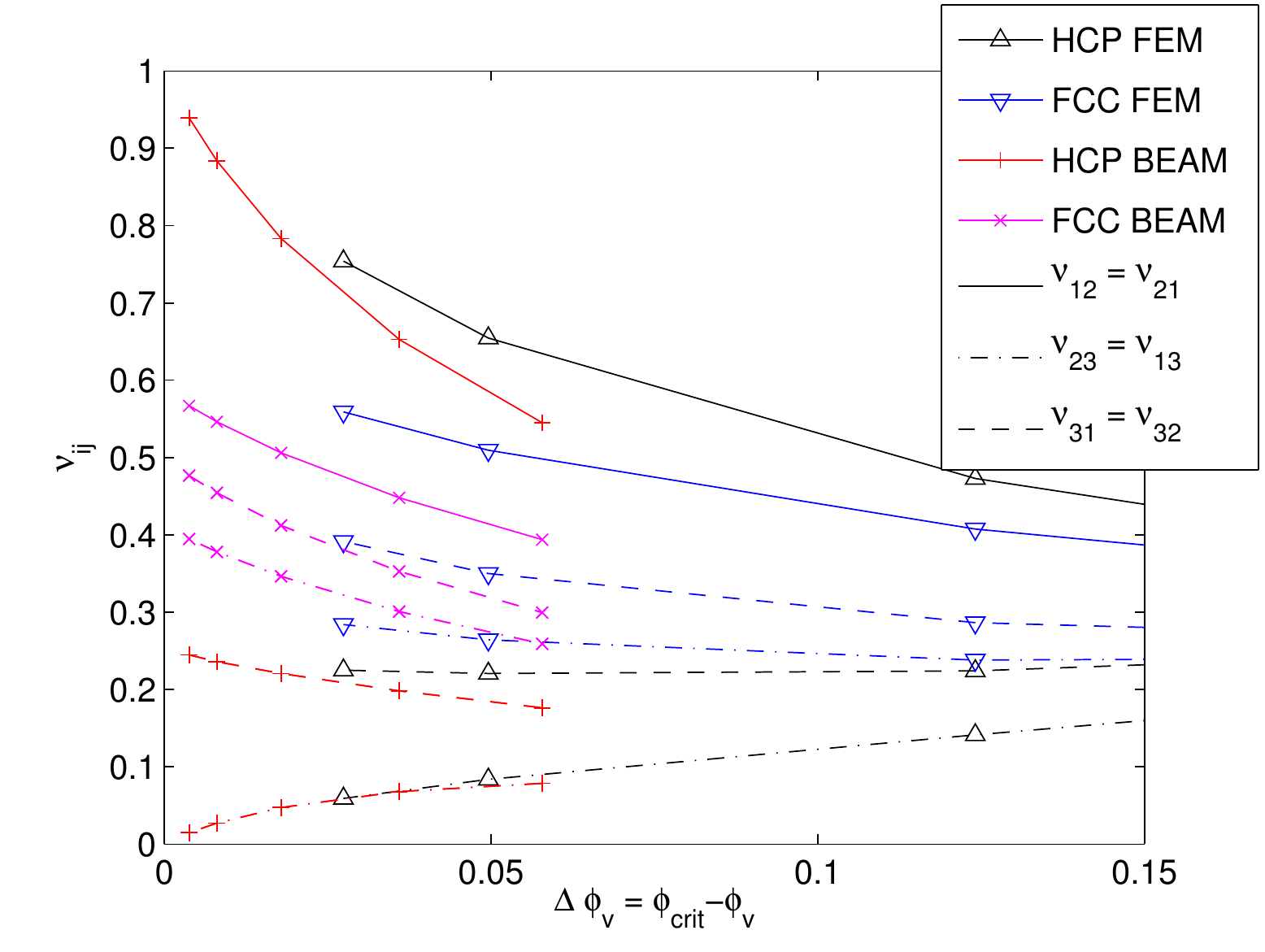}
\caption{Poisson ratio as a function of the difference between void fraction and critical void fraction. Data of FE simulations ($\nabla$, $\Delta$) and Beam model ($\times$,+) for FCC ($\nabla$, $\times$) and HCP ($\Delta$,+) for different combinations of directions $e_i,e_j$. Here, $e_3$ corresponds to the symmetric axis; $e_1$ and $e_2$ correspond to directions in the basal plane. Due to isotropy in the basal plane, some combinations of direction coincide.}
\label{fig:nu}       
\end{figure}

\begin{figure}
\centering
  \includegraphics[width=0.99\textwidth]{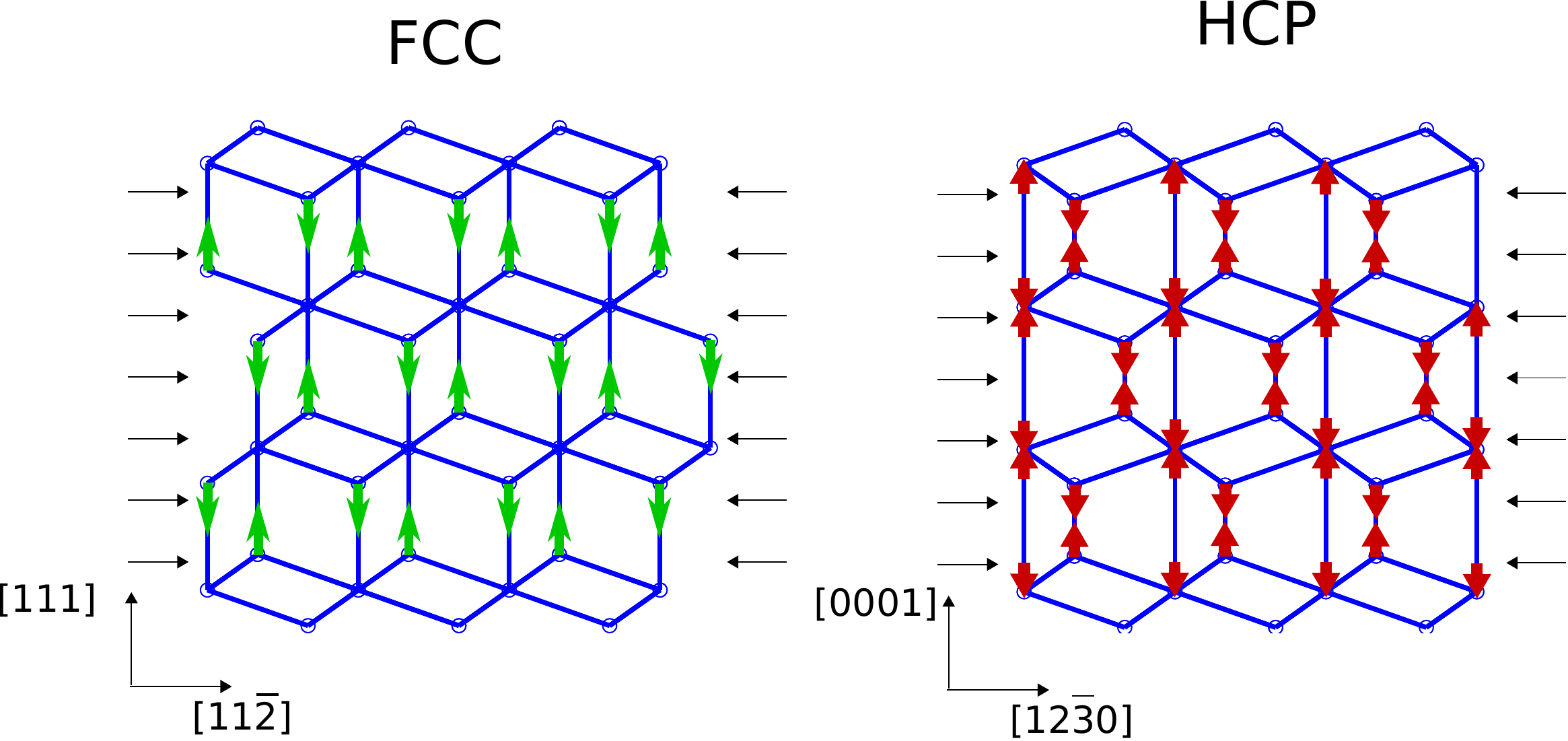}
\caption{Appearance of deformations along the symmetry axis when compressing an FCC or HCP void material in a direction parallel to the basal plane. Horizontal arrows mark the applied strain. Vertical arrows mark the resulting shift of vertices. FCC topology allows for easy shear of the network while HCP topology generates internal stress but no vertical strain.}
\label{fig:network2}       
\end{figure}

The Poisson ratios predicted by the beam model also show good agreement with the FE data as shown in Figure~\ref{fig:nu}. The relations between different Poisson ratios and also the trend for increasing void fraction are well reproduced. 
Remarkable is the fact, that some values of $\nu_{12}$ are larger than 0.5. This would be impossible for homogeneous material, but for void material this is possible~\cite{Ting2005}. Especially for $\nu_{12}$ in HCP, Poisson ratios up to $1$ appear. This means, that a compression in the basal plane, along the \hkl [1 0 -1 0] direction causes an equal expansion in the basal plane, along the \hkl [1 2 -3 0] direction. At the same time, $\nu_{13}$ is close to zero, so nearly no expansion along the symmetric axis appears. For FCC, in contrast, $\nu_{12}$ does not exceed 0.7, and also $\nu_{13}$ does not fall below 0.25. This is remarkable, because both structures consist of the same type of material layer in the basal plane, but due to the different orientations and interconnections of these layers, different elastic behaviour occurs. This situation is shown in Figure~\ref{fig:network2}.\\
When compressing a material layer in a direction parallel to the basal plane, the small vertices $\beta$ are pushed outwards in the direction of the symmetric axis. In case of FCC, they are connected with a large vertex $\alpha$ in the next layer, so that they push away the next layer resulting in a moderate Poisson ratio. Additionally, a small shear deformation appears. This is expressed in~\cite{Heitkam2015} by non-ortotrophy of the FCC void material. In contrast, $\beta$ connects with $\beta$ and $\alpha$ connects with $\alpha$ in the HCP arrangement. When compressing a material layer, the $\beta$ vertices are then pushed against each other. This would cause a large expansion in the symmetric direction. At the same time, links between $\alpha$ vertices counteract the expansion, resulting in an overall small Poisson ratio $\nu_{13}$. Instead, the structure relaxes in the basal plane, yielding values close to 1 for $\nu_{12}$. When the void fraction approaches the critical void fraction, stretching becomes increasingly stiffer than bending. Consequently, $\nu_{13}$ further decreases and $\nu_{12}$ further increases.

\section{Discussion and conclusions}
In general, the simple beam model is able to reproduce the predictions from the FE simulations and to extend it to lower void fractions which cannot be reached by FE simulations. In an early version of the beam model used by the authors, beams with constant thickness were employed. This also reproduced the general relations between different structures and directions, but was not able to represent the power laws linking void fraction and elastic properties. The present version is able to do so. Consequently, the distribution of material along the beams plays an important role for the elastic properties. This also indicates some kind of limitation to the present results, because for realistic foams the mass is more equally distributed along the beams~\cite{Lambert2005}. In a later version of the beam model, one could extract elastic constants from x-ray tomography~\cite{Jang2008} or theory. For materials generated from 3D-printing or sacrificial templating the present material distribution is presumably more realistic. 

Several studies find for bending-dominated open-cell beam networks a dependence of the Young's modulus $E \sim (1-\phi_v)^{2}$~\cite{Gibson1997,Warren1997,Roberts2002}. This finding is due to the constant thickness $t$ of a beam along its length employed in those studies. In those cases and for constant beam length, the void fraction scales with $ (1-\phi_v)  \sim t^2 $ and the Young's modulus with $E \sim I_p \sim t^4$. 
In the present study $E \sim (\Delta \phi_v)^{5/2}$ was found. The difference in the exponent results from different material distribution along the beam. Most of the material is concentrated in the nodes causing a weaker dependence of the beam thickness on the void fraction $h_0 \sim (\Delta \phi_v)^{1}$. Also the Young's modulus as $E \sim k_{bend} \sim h_0^{5/2}$ in the present case is more sensitive to the void fraction than in case of constant beam cross section. In conclusion the scaling of constant cross section and our model compares as follows:
\begin{eqnarray}
 \mathrm{Constant \; cross\;  section: }& \;(1-\phi_v)  \sim t^2 \; , \;  E \sim I_p \sim t^4 \; , \; &\rightarrow \;  E \sim (1-\phi_v)^{2} \nonumber \\
 \mathrm{Our \;beam \; model: }& \; (\Delta \phi_v)  \sim h_0^1 \; , \;  E \sim k_{bend} \sim h_0^{5/2} \; , \; & \rightarrow \;  E \sim (\Delta \phi_v)^{5/2} \nonumber 
\end{eqnarray}
For HCP void arrangement, in \hkl [0 0 0 1] direction, a different dependence of the Young's modulus on the void fraction exhibits. In this direction, the interstitial material on HCP void arrangement shows a straight, unbroken fibre of connected beams (Figure~\ref{fig:void_material}c). It connects vertices of type $\alpha$. This fibre causes a stretching-dominated elastic behaviour, as can be derived from symmetry arguments: The material layers, marked in Figure~\ref{fig:void_material}c, show 3-fold symmetry and thus, isotropy in the basal plane. 
Due to this isotropy, forces on the $\alpha$-vertices in the basal plane balance each other (buckling is not considered here) and the $\alpha$-vertices are not shifted parallel to the basal plane when applying strain normal to the basal plane.   
Consequently, only shifting of the $\alpha$ vertices along the symmetric axis is allowed, which corresponds to stretching of the fibre. At the same time, the Poisson ratios $\nu_{[0001],[10\bar{1}0]}$ and $\nu_{[0001],[12\bar{3}0]}$ go to zero in the limit of vanishing beam thickness. Stretching of the fibres corresponds to stretching of necks. The corresponding elastic parameter of the beams, $k_{stretch} \sim (\Delta \phi_v)^{3/2}$ is more sensitive to the void fraction. Again, comparison to a model with constant beam thickness $t$ reads as follows:
\begin{eqnarray}
 \mathrm{Constant \; cross\;  section: }& \;(1-\phi_v)  \sim t^2 \; , \;  E \sim A_s \sim t^2 \; , \; &\rightarrow \;  E \sim (1-\phi_v)^{1}\nonumber  \\
 \mathrm{Our \;beam \; model: }& \; (\Delta \phi_v)  \sim h_0^1 \; , \;  E \sim k_{stretch} \sim h_0^{3/2} \; , \; & \rightarrow \;  E \sim (\Delta \phi_v)^{3/2} \nonumber 
\end{eqnarray}
Discussing the FE results with colleagues in the foam community raised scepticism about Poisson ratios higher than 0.5, since this is rigorously excluded for isotropic elastic materials. However, in the general case the Poisson ratio has no bounds~\cite{Ting2005}. The beam model now provides a demonstration of the appearance of very high or very low values. The Poisson ratio of the solid material was set to 0.4 in order to represent the case of polymers, but in fact, the influence of the material Poisson ratio on the results at high void fraction is negligible. In other structured materials these extreme Poisson ratios are well-known, including artificial materials with negative Poisson ratios by molecular design~\cite{He1998} or by structure~\cite{Lee2015}. In the field of fibre-reinforced materials, very high ratios of longitudinal and transversal Young's modulus and Poisson ratios are utilised~\cite{Bakis2002,Mallick2007}. This kind of anisotropic behaviour enhances the fracture resistance~\cite{Kim1991} and vibration damping~\cite{Chandra1999}. Open-cell HCP void material shows strongly anisotropic behaviour because of straight fibres in the \hkl [0 0 0 1] direction. To benefit from these, one needs to create open-cell HCP void material. This could be done with sacrificial templating of soft spheres for example. Under confinement and with vibration~\cite{Yu2006} they will arrange in dense-packed lattices at a void fraction of $0.74~\%$. Compression could increase the void fraction. Unfortunately, spheres prefer FCC and not HCP arrangement when packed~\cite{Heitkam2012}. Therefore, additional templating techniques need to be applied to promote HCP arrangement.\\
In future work, the beam model may be applied to different void structures such as simple cubic, body-centred cubic or Weaire-Phelan structure~\cite{Weaire1994} or even random structure. Including representation of lamellas in the beam model would allow for estimation of closed-cell materials with finite lamella thickness. The beam model can provide an efficient and complement model for design and analysis of innovative void materials. The more demanding full FE simulations are necessary only when the void fraction is relatively small.


\section*{Acknowledgements}
We gratefully acknowledge Frederic Piechon, Christophe Poulard, Thomas Titscher, Daniel Christopher Kreuter, and David Hajnal for fruitful discussions on the topic and during the work that originally motivated the present study~\cite{Heitkam2015}. Computation time was provided by the Center for Information Services and High Performance Computing (ZIH) at TU Dresden. We acknowledge support from the European Research Council (ERC) under the European Union’s Seventh Framework Program (FP7/2007-2013) in form of an ERC Starting Grant, agreement 307280-POMCAPS. We acknowledge support from the European Centre for Emerging Materials and Processes (ECEMP) at TU Dresden and the Helmholtz-Alliance Liquid Metal Technologies (LIMTECH).


\begin{thebibliography}{10}
\providecommand{\url}[1]{{#1}}
\providecommand{\urlprefix}{URL }
\expandafter\ifx\csname urlstyle\endcsname\relax
  \providecommand{\doi}[1]{DOI~\discretionary{}{}{}#1}\else
  \providecommand{\doi}{DOI~\discretionary{}{}{}\begingroup
  \urlstyle{rm}\Url}\fi

\bibitem{Bakis2002}
Bakis, C., Bank, L.C., Brown, V., Cosenza, E., Davalos, J., Lesko, J., Machida,
  A., Rizkalla, S., Triantafillou, T.: Fiber-reinforced polymer composites for
  construction-state-of-the-art review.
\newblock Journal of Composites for Construction \textbf{6}(2), 73--87 (2002)

\bibitem{Chandra1999}
Chandra, R., Singh, S., Gupta, K.: Damping studies in fiber-reinforced
  composites--a review.
\newblock Composite structures \textbf{46}(1), 41--51 (1999)

\bibitem{Day1992}
Day, A., Snyder, K., Garboczi, E., Thorpe, M.: The elastic moduli of a sheet
  containing circular holes.
\newblock Journal of the Mechanics and Physics of Solids \textbf{40}(5), 1031
  -- 1051 (1992)

\bibitem{Gibson1997}
Gibson, L.J., Ashby, M.F.: Cellular solids: structure and properties.
\newblock Cambridge University Press, 2nd Edition (1997)

\bibitem{He1998}
He, C., Liu, P., Griffin, A.C.: Toward negative poisson ratio polymers through
  molecular design.
\newblock Macromolecules \textbf{31}(9), 3145--3147 (1998)

\bibitem{Heitkam2012}
Heitkam, S., Drenckhan, W., Fr\"ohlich, J.: Packing spheres tightly: Influence
  of mechanical stability on close-packed sphere structures.
\newblock Phys. Rev. Lett. \textbf{108}, 148,302 (2012)

\bibitem{Heitkam2015}
Heitkam, S., Drenckhan, W., Titscher, T., Kreuter, D., Hajnal, D., Piechon, F.,
  Fr\"ohlich, J.: Elastic properties of material with spherical voids in
  different arrangements.
\newblock Submitted  (2015)

\bibitem{Heitkam2013}
Heitkam, S., Schwarz, S., Santarelli, C., Fr{\"o}hlich, J.: Influence of an
  electromagnetic field on the formation of wet metal foam.
\newblock The European Physical Journal Special Topics \textbf{220}(1),
  207--214 (2013)

\bibitem{Jang2008}
Jang, W.Y., Kraynik, A.M., Kyriakides, S.: On the microstructure of open-cell
  foams and its effect on elastic properties.
\newblock International Journal of Solids and Structures \textbf{45}(7),
  1845--1875 (2008)

\bibitem{Kikuchi2011}
Kikuchi, K., Ikeda, K., Okayasu, R., Takagi, K., Kawasaki, A.: Structural
  dependency of three-dimensionally periodic porous materials on elastic
  properties.
\newblock Materials Science and Engineering: A \textbf{528}(28), 8292--8298
  (2011)

\bibitem{Kim1991}
Kim, J.K., Mai, Y.w.: High strength, high fracture toughness fibre composites
  with interface controla review.
\newblock Composites Science and Technology \textbf{41}(4), 333--378 (1991)

\bibitem{Lambert2005}
Lambert, J., Cantat, I., Delannay, R., Renault, A., Graner, F., Glazier, J.A.,
  Veretennikov, I., Cloetens, P.: Extraction of relevant physical parameters
  from 3d images of foams obtained by x-ray tomography.
\newblock Colloids and Surfaces A: Physicochemical and Engineering Aspects
  \textbf{263}(1), 295--302 (2005)

\bibitem{Lee2015}
Lee, J., Kim, K., Ju, J., Kim, D.M.: Compliant cellular materials with
  elliptical holes for extremely high positive and negative poisson's ratios.
\newblock Journal of Engineering Materials and Technology \textbf{137}(1),
  011,001 (2015)

\bibitem{Mallick2007}
Mallick, P.K.: Fiber-reinforced composites: materials, manufacturing, and
  design.
\newblock CRC press (2007)

\bibitem{Mbakogu1987}
Mbakogu, F., Pavlovi{\'c}, M.: Shallow shells under arbitrary loading: Analysis
  by the two-surface truss model.
\newblock Communications in applied numerical methods \textbf{3}(3), 235--242
  (1987)

\bibitem{Mills2007}
Mills, N.: Polymer foams handbook: engineering and biomechanics applications
  and design guide.
\newblock Butterworth-Heinemann (2007)

\bibitem{Roberts2002}
Roberts, A., Garboczi, E.J.: Elastic properties of model random
  three-dimensional open-cell solids.
\newblock Journal of the Mechanics and Physics of Solids \textbf{50}(1), 33--55
  (2002)

\bibitem{Salonen1971}
Salonen, E.M.: Triangular framework model for plate bending.
\newblock Journal of the Engineering Mechanics Division \textbf{97}(1),
  149--153 (1971)

\bibitem{Ting2005}
Ting, T., Chen, T.: Poisson's ratio for anisotropic elastic materials can have
  no bounds.
\newblock The quarterly journal of mechanics and applied mathematics
  \textbf{58}(1), 73--82 (2005)

\bibitem{Warren1997}
Warren, W., Kraynik, A.: Linear elastic behavior of a low-density kelvin foam
  with open cells.
\newblock Journal of Applied Mechanics \textbf{64}(4), 787--794 (1997)

\bibitem{Weaire2008}
Weaire, D., Aste, T.: The pursuit of perfect packing.
\newblock CRC Press (2008)

\bibitem{Weaire1994}
Weaire, D., Phelan, R.: A counter-example to kelvin's conjecture on minimal
  surfaces.
\newblock Philosophical Magazine Letters \textbf{69}(2), 107--110 (1994)

\bibitem{Yu2006}
Yu, A., An, X., Zou, R., Yang, R., Kendall, K.: Self-assembly of particles for
  densest packing by mechanical vibration.
\newblock Physical review letters \textbf{97}(26), 265,501 (2006)

\end{thebibliography}
\end{document}